\documentclass[graybox]{svmult}

\usepackage{mathptmx}       
\usepackage{helvet}         
\usepackage{courier}        
\usepackage{type1cm}        
\usepackage{makeidx}         
\usepackage{graphicx}        
\usepackage{multicol}        
\usepackage[bottom]{footmisc}
\usepackage{textcomp}
\usepackage{gensymb}
\usepackage{natbib}

\makeindex             

\begin{document}

\title*{Nuclear Star Clusters and Bulges}
\author{David R. Cole, Victor P. Debattista}
\institute{David R. Cole, Jeremiah Horrocks
  Institute, University of Central Lancashire, Preston PR1 2HE,
  \email{drdrcole@gmail.com}, \& Victor P. Debattista, Jeremiah Horrocks Institute,
  University of Central Lancashire, Preston PR1 2HE,
  \email{vpdebattista@gmail.com} }
%
%
\maketitle

\abstract{Nuclear star clusters are among the densest stellar systems
  known and are common in both early- and late-type galaxies. They
  exhibit scaling relations with their host galaxy which may be
  related to those of supermassive black holes.  These may therefore
  help us to unravel the complex physical processes occurring at the
  centres of galaxies. The properties of nuclear stellar systems
  suggest that their formation requires both dissipational and
  dissipationless processes. They have stellar populations of
  different ages, from stars as old as their host galaxy to young
  stars formed in the last 100 Myr. Therefore star formation must be
  happening either directly in the nuclear star cluster or in its
  vicinity.  The secular processes that fuel the formation of
  pseudobulges very likely also contributes to nuclear star cluster
  growth. }

\section{Introduction}
\label{sec:intro}

Observations with the high resolution instruments on the {\it Hubble
  Space Telescope} ({\it HST}) have revealed that many low to
intermediate mass galaxies contain a dense stellar system at their
centre.  They are among the densest stellar systems known.  Nuclear
stellar systems come in two main morphological types, nuclear star
clusters (NSCs), where the stellar distribution is spheroidal, and
nuclear discs (NDs).  These are not mutually exclusive and NSCs often
contain a disc component too, typically comprised of younger stars.
These dense stellar systems are common in galaxies across the Hubble
sequence.

A connection between NSCs and the formation of their host galaxy is
implied by various observed scaling relations between their mass and
the properties of their host. These scaling relations provide insight
into the physical processes regulating the growth of nuclear stellar
systems.

NSCs in late-type disc galaxies are observed to have a mix of
populations, including young stars formed within the last 100 Myr.
Whether these formed in situ, or arrived as star clusters accreted
from the neighbourhood of the NSC, gas is needed to make these young
stars.  Thus NSCs and NDs provide evidence that gas is able to
repeatedly reach the centres of late-type galaxies.

\section{Nuclear Stellar Systems}
\label{sec:Nss}

\subsection{Properties and occurrence}
\label{subsec:nsprop}

NSCs are compact objects with effective radii of order 5 pc and masses
ranging from $10^5 M_\odot$ to $10^8 M_\odot$, meaning they have among
the highest known average surface densities \citep{Walcher2005}.  NSCs
have photometric and kinematic properties broadly similar to those of
globular clusters but with higher velocity dispersions.
Their absolute visual magnitudes lie between -14 and -10
\citep{Boker2002, Cote2006} compared with Milky Way globular clusters
which have absolute magnitudes typically in the range -9 to -4
\citep{Harris1997}.  \cite{Hartmann2011} found that the NSC in M33 was
photometrically and kinematically consistent with being perfectly
axisymmetric.

NSCs are present in between $50\%$ and $75\%$ of low to intermediate
luminosity galaxies. \cite{Carollo1997} found NSCs in 18 of 35
\textit{HST} WFPC2 F606W images of spiral galaxies including
early-types, while \cite{Boker2002} found NSCs in 59 of 77
\textit{HST} images of late-type spiral galaxies.  Between $66\%$ and
$82\%$ of early-type galaxies in the \textit{HST} ACS Virgo Cluster
Survey have NSCs \citep{Cote2006}, and a similar fraction in the ACS
Fornax Cluster Survey \citep{Turner2012}.  The presence of a bar does
not seem to affect whether a NSCs occurs or not \citep{Carollo2002,
  Boker2004}.

Nuclear discs are also often found in the central regions of galaxies.
They span a range of sizes from a few parsecs to of order a kiloparsec
in diameter.  They can be differentiated from the main galactic disc
(if it exists) in that they lie outside of the region where light from
the main disc dominates. They are widely observed in galaxies spanning
the full range of Hubble types both in late-type galaxies
\citep{Zasov1999, Pizzella2002, Dumas2007, GarciaBurillo2012} and in
early-types \citep{Scorza1998, Kormendy2001, deZeeuw2002, Emsellem2004,
  Trujillo2004, Krajnovic2008, Ledo2010}.  \cite{Ledo2010} found that
as many as $20\%$ of early-type galaxies host a nuclear disc.  A
sample of 48 early-type galaxies observed as part of the SAURON
project, revealed that nuclear discs are associated with early-type
fast rotators \citep{Krajnovic2008}.

\subsection{Stellar ages}
\label{subsec:nsages}

NSCs in late-type galaxies often consist of multiple stellar
populations. Their mean luminosity-weighted ages range from 10 Myr to
10 Gyr \citep{Rossa2006}, very often with evidence of star formation in
the last 100 Myr \citep{Walcher2005, Walcher2006}.  Spectra reveal that
their star formation is bursty, with a duty cycle of a few hundred
Myr.  For instance, the NSC in M33 had bursts of star formation 40 Myr
and 1 Gyr ago \citep{Long2002}. \cite{Georgiev2014}, in a study of
  NSCs in 228 late-type galaxies, also find their stellar populations
  span a wide range of ages and conclude that recent star formation is
  ubiquitous.

\begin{figure}[b]
\sidecaption
\includegraphics[scale=0.6]{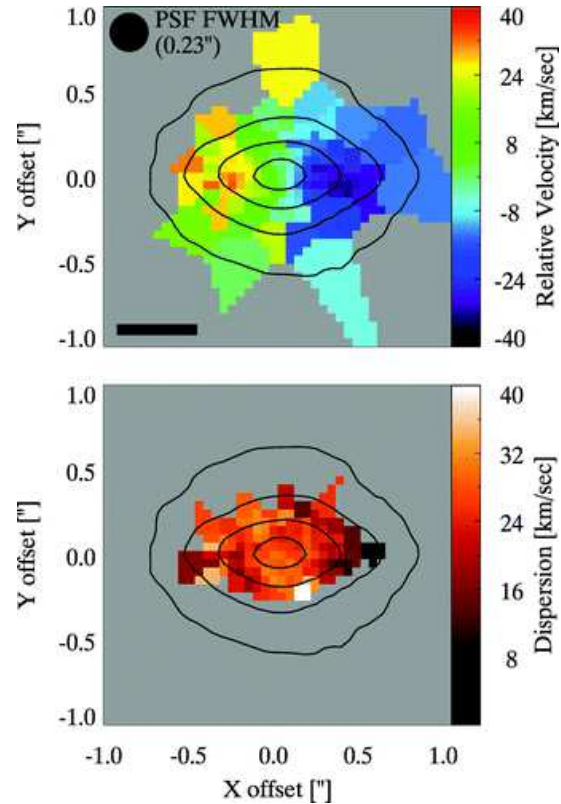}
\caption{The kinematics of the NSC in NGC~4244. Top: The measured
  radial velocity observed with NIFS. Rotation of 30 km s$^{-1}$ is
  clearly visible along the major axis. Contours show the K-band
  isophotes. The black bar indicates 10 pc (0.47$^{\prime \prime}$ ).
  Bottom: Velocity dispersion measurements. 
  Figure 2 of \cite{Seth2008}.}
\label{fig:S08} 
\end{figure}

NSCs in late-type disc galaxies are typically elongated approximately
in the plane of the main galaxy disc and are often made up of two
components, an older spheroidal component, with a younger and bluer
disc embedded in it \citep{Seth2006, Seth2008}. In the case of
NGC~4244, spectra indicate young ($<100$ Myr) stars in its disc.
Integral field spectroscopy reveals that the NSC has rotation in the
same sense as the galaxy (see Figure \ref{fig:S08}) with a relative
tilt of $15\degree$. Similarly the NSC of FCC~277, an elliptical
galaxy in the Fornax cluster, is made up of a spheroid and a disc
component, both of which are younger than the main galaxy
\citep{Lyubenova2013}.  \cite{Carson2015} studied {\it HST} WFC
  images of 10 of the nearest and brightest NSCs. They found
  increasing roundness at longer wavelengths inferring the existence
  of blue discs made up of younger stellar populations as in NGC~4244.
  Most of these NSCs show evidence in colour-colour diagrams of
  stellar populations consisting of a mixture of an older population
  ($>$1 Gyr) and a younger population (100 - 300 Myr).
\cite{Pfuhl2011} estimated that the NSC in the Milky Way formed $\sim
80\%$ of its stars more than 5 Gyr ago, with a deep minimum in star
formation 1 to 2 Gyr ago.  The star formation rate then increased
again in the last few hundred Myr.

Likewise NDs often exhibit a range of ages with a tendency for young
stars to be present.  Ongoing star formation is observed in the NDs of
NGC~5845 \citep{Kormendy1994} and NGC~4486A \citep{Kormendy2005}.  In
NGC~4486A the ND manages to have stars more than 2 Gyr younger than
the surrounding galaxy \citep{Kormendy2005}.  The ND in NGC~4570 shows
evidence for recent star formation \citep{vandenBosch1998}.
\cite{Morelli2004} found that NGC~4478 has a younger stellar
population than the main body of the galaxy, with a prolonged star
formation history, whereas NGC~4458 has a uniformly old population.
The ND stellar population in NGC~4698 has ages in the range 5 to 10
Gyr \citep{Corsini2012}.  On the other hand, NDs in early-type galaxies
have been found to consist mainly of old ($ > 10$ Gyr) stars
\citep{Krajnovic2004}.

\subsection{Kinematic decoupling}
\label{subsec:prop}

One phenomenon which suggests that interactions may play a role in the
formation of nuclear stellar systems is kinematic decoupling where
distinct stellar components have large ($\geq40\degree$) misalignments
in their axes of rotation \citep{McDermid2006}. About one third of
early-type galaxies in the SAURON sample exhibit this decoupling.
Kinematic decoupling has been interpreted as evidence for formation
through the capture of external gas. Two types are observed, one on
kiloparsec scales, which are generally older than 8 Gyr and found in
galaxies with little net rotation, and one which consists of
structures on the scale of a few hundred parsec and which have ages
ranging from 500 Myr to a Hubble time \citep{McDermid2006}.  The ND in
NGC~4458 is counter-rotating implying the gas had an external origin
\citep{Morelli2004, Morelli2010}. NGC~4698 also displays kinematical
decoupling and has a disc rotating perpendicular to the main galactic
disc suggesting that the ND formed from externally accreted gas
\citep{Bertola1999, Pizzella2002, Corsini1999, Corsini2012}.
While the presence of kinematically decoupled cores demonstrates
  that externally captured gas can reach small radii, this by no means
  implies that the gas also forms a nuclear cluster.  For instance,
  although NGC~4458 has a stellar disc it has been classified as
  non-nucleated \citep{Lauer2005}.

\section{Formation}
\label{sec:nssform}

\subsection{Nuclear Star Clusters}
\label{subsec:nsform}

Two principal formation mechanisms have been advanced to explain the
formation of NSCs. The first is that NSCs form due to globular
clusters falling to the centres of galaxies under the action of
dynamical friction and subsequently merge \citep{Tremaine1975,
  CapuzzoDolcetta1993, Miocchi2006, CapuzzoDolcetta2008a,
  CapuzzoDolcetta2008b, Antonini2012, Antonini2013, Gnedin2014}.  A
candidate infalling globular cluster was found in the inner few
hundred parsecs of NGC~2139 which could become a NSC in a few hundred
Myr \citep{Andersen2008}.  An off-centre super star cluster with a mass
of $1.4^{+0.4}_{-0.5}\times10^7$ M$_\odot$ has been observed in
NGC~253.  This super star cluster is a candidate future NSC
\citep{Kornei2009}.  \cite{Georgiev2014} present NGC~4654 as an example
of galaxies where two star clusters are present at the centre. The two
star clusters have a mass ratio of order $10:1$ and are separated by
$\sim 30$ pc (in projection).  The less massive of the star clusters
appears to be young ($< 100$ Myr), supporting the picture of NSC
growth due to the accretion of {\it young} globular clusters onto the
centre. \cite{Nguyen2014} find a starburst at the centre of a
SMBH-hosting galaxy Heinze 2-10. This starburst has created several
super star clusters within 100 pc of the SMBH and they conclude that
the star clusters would merge due to dynamical friction and form an
NSC in the next Gyr. Simulations by \cite{CapuzzoDolcetta1993} showed
that, while the infall of globular clusters is in competition with
tidal stripping (which destroys the clusters), nonetheless a fraction
of them do manage to reach the nucleus.  Fewer massive globular
clusters than expected have been found in the inner region of dwarf
ellipticals suggesting that the globular clusters missing from the
inner regions had been depleted due to their shorter dynamical
friction infall times and merging to form NSCs \citep{Lotz2001,
  Lotz2004}.  The infall of globular clusters has been modelled and
simulated several times.  \cite{Agarwal2011} modelled the infall (and
stripping) of globular clusters analytically and concluded that this
process could create NSCs which match the observed NSC masses.
\cite{CapuzzoDolcetta2008b} show that globular clusters merging at the
centre of a bulge leads to density-velocity dispersion properties
consistent with those of observed NSCs.  Analytic modelling of the
infall of globular clusters led \cite{Gnedin2014} to the same
conclusion; they argue, moreover, that the contrast between the NSC
and the background galaxy would be much lower in a massive galaxy like
M87, making them harder to detect in such systems.  \cite{DenBrok2014}
studied NSCs in 200 Coma cluster dwarf elliptical galaxies and found a
relation between NSC and host galaxy magnitude of $M_{nuc} =
(0.57\pm0.05)(M_{gal}+17.5)-(11.49\pm0.14)$ concluding that this is
consistent with predictions of how NSC luminosity scales with host
galaxy luminosity as predicted by the globular cluster merger scenario
models of \cite{Antonini2013} and \cite{Gnedin2014}. However they also
find that galaxies with higher S\'ersic indices tend to have brighter
NSCs. They argue that this is due to the brighter galaxies being
better able to retain gas and conclude that in situ star formation
also plays an important role.

\begin{figure}[b]
\sidecaption
\includegraphics[scale=0.6]{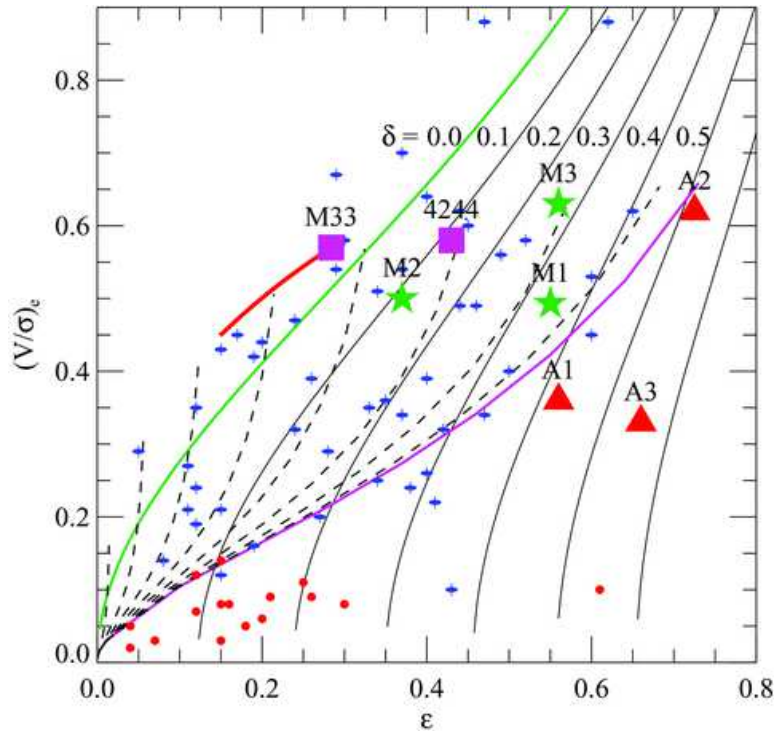}
\caption{Real and simulated NSCs on the (V$/\sigma,\epsilon$) diagram
  of \cite{Binney2005}. The green stars and red triangles indicate the
  location of the simulated NSCs (M1-M3 \& A1-A3) formed by merging
  infalling globular clusters, while the magenta squares are the
  observed NSCs in NGC~4244 and M33. The observed location for M33 has
  been projected (red line) assuming an inclination of$ i =
  49\degree$. For comparison, the blue symbols with vertical axes and
  red filled circles are the observed fast and slow rotator early-type
  galaxies, respectively, from \cite{Cappellari2007}. Figure 5 of
  \cite{Hartmann2011}.}
\label{fig:H11} 
\end{figure}

\cite{DenBrok2014} showed an example of an unnucleated galaxy with
several old globular clusters within it which raises the question of
how this galaxy managed to avoid forming a NSC.  \cite{Hartmann2011}
note that the NSC in M33 appears axisymmetric (both photometrically
and kinematically); their simulations of globular cluster mergers
produce triaxial NSCs although this outcome can be avoided if a
massive black hole is present.  However in M33 the upper limit on the
presence of a black hole is very stringent (M$_{\rm bh} < 3000
M_\odot$, \citep{Merritt2001, Gebhardt2001}) suggesting that its NSC
did not form via globular cluster mergers. On the other hand, the NSC
in the Milky Way has a rotating sub-structure perpendicular to the
Galactic plane \citep{Feldmeier2014}, suggesting a cluster was
accreted. in situ star formation has been proposed as an alternative
for forming NSCs \citep{Milosavljevic2004, Bekki2007}.  \cite{Cen2001}
suggested that at the epoch of re-ionisation the external radiation
field could create an inward convergent shock leading to the formation
of massive dense clusters at the centres of early galaxies with masses
and velocity dispersions comparable to those of NSCs.  While these
could have formed the seeds of some NSCs, the NSCs would have had to
grow further since formation to account for the younger populations.
\cite{Emsellem2008} showed that the tidal field of a wide range of
S\'{e}rsic profile spheroids are compressive in the regions where NSCs
form; gas falling in is therefore likely to form stars.  They found
that the mass of the object expected to form would be $0.1-0.5\%$ that
of the host, consistent with the masses of both SMBHs and NSCs.  The
most direct evidence for the need of in situ star formation comes from
modelling the kinematics of the NSC in NGC~4244 (Figure
\ref{fig:S08}).  Simulations by \cite{Hartmann2011} find that though
the globular cluster merging scenario can reproduce many of the
density and kinematic properties of NSCs (see Figure \ref{fig:H11}),
mergers give rise to a central peak in $v_{\rm rms} =
\sqrt{\sigma_{\rm los}^2 + v_{\rm los}^2}$, which is not observed in
the data.  Based on this, they conclude that less than $50\%$ of the
mass of the NSC could have been assembled from the mergers of globular
clusters, with the majority due to in situ star formation.  On the
other hand, they also find a negative vertical anisotropy, $\beta_z =
1 - \sigma_z^2/\sigma_R^2$, confirmed through the independent
modelling of \cite{DeLorenzi2013} using the made-to-measure technique
\citep{Syer1996, DeLorenzi2007}.  This, they showed, could be produced
by the accretion of a globular cluster, accounting for at least $10\%$
of its mass, on a nearly polar orbit relative to the NSC.  They
conclude that both in situ star formation and globular clusters
mergers played a role in the formation of this NSC.  From a sample of
over 200 late-type spiral galaxies observed with {\it HST},
\cite{Georgiev2014} showed that NSCs are smaller in blue compared to
red filters.  This can be explained either by the presence of an AGN
or by population gradients within the NSC, possibly indicating ongoing
star formation.  \cite{Turner2012} studied the nuclei in 43 early-type
galaxies in the Fornax cluster.  On the basis of globular cluster
infall times, they concluded that in low mass early-type galaxies the
dominant mechanism for NSC formation is probably globular cluster
merging but for more massive galaxies in situ star formation becomes
necessary.  A picture is being established therefore where both
processes, globular cluster mergers and in situ star formation, play a
role in NSC formation and which is the dominant mechanism depends on
the parameters of the host galaxy \citep{Rossa2006, Walcher2006}.

\subsection{Nuclear Disc Formation}
\label{subsec:ndformation}

\begin{figure}[b]
\sidecaption
\includegraphics[scale=1.0]{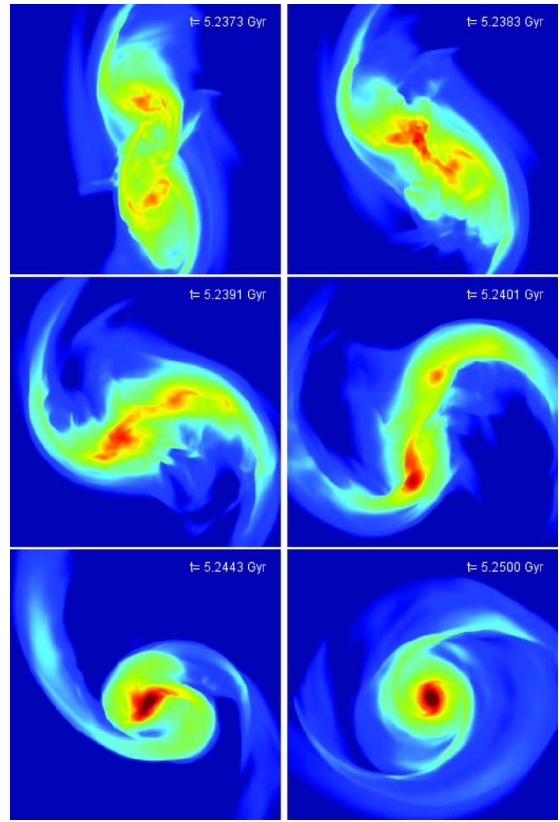}
\caption{ Gas density maps during the final stages of the merger of
  two galaxies. We are looking down onto the orbital plane of the
  galaxies and the maps are 1.8 kpc wide. Gas from the two galaxies
  funnels inwards to form a thick gaseous nuclear disc with two SMBHs
  orbiting in it in the final image. Figure 1 of \cite{Chapon2013}}
\label{fig:galmerg} 
\end{figure}

  The formation of NDs is thought to require in situ star
formation.  A significant mass of gas needs to be funnelled to the
nuclear regions to allow this. Such inflows are possible in mergers as
shown by hydrodynamical simulations \citep{Mayer2008, Mayer2010,
  Hopkins2010, Chapon2013}.  \cite{Chapon2013} presented a simulation
of the merger of two galaxies with SMBHs (see Figure
\ref{fig:galmerg}).  After the merger a thick nuclear gas disc forms
with a mass $\sim$ 10$^9$ M$_\odot$. Nuclear discs have been observed
in 17 nearby luminous infra-red galaxies (LIRGs) and ultra-luminous
infra-red galaxies (ULIRGs), possibly the results of merger-driven gas
funnelling to their centres initiating intense star formation
\citep{Medling2014}.  Meanwhile \cite{Hopkins2010} showed that lopsided
nuclear discs such as the one in the Andromeda galaxy may form via
merger-driven inflows.  Instead \cite{Ledo2010} showed that
pre-existing NDs are destroyed in mergers.

Secular processes such as the action of a bar can also supply gas to
form a ND. The formation of the ND in the edge-on galaxy NGC~7332 was
attributed to the presence of a bar (which was inferred from the
boxy/peanut-shaped bulge) \citep{Seifert1996, FalconBarroso2004}.
Barred galaxies have more molecular gas in their central kiloparsec
than unbarred galaxies \citep{Sakamoto1999, Sheth2005}.  Enhanced
nuclear star formation correlates with the presence of a strong bar in
disc galaxies \citep{Wang2012}, and depends primarily on the
ellipticity of the bar, not on the size of the bar.  However only half
of galaxies with centrally concentrated star formation have a strong
bar suggesting that processes such as interactions with other galaxies
also induce star formation in the nucleus.

\begin{figure}[b]
\sidecaption
\includegraphics[scale=0.4]{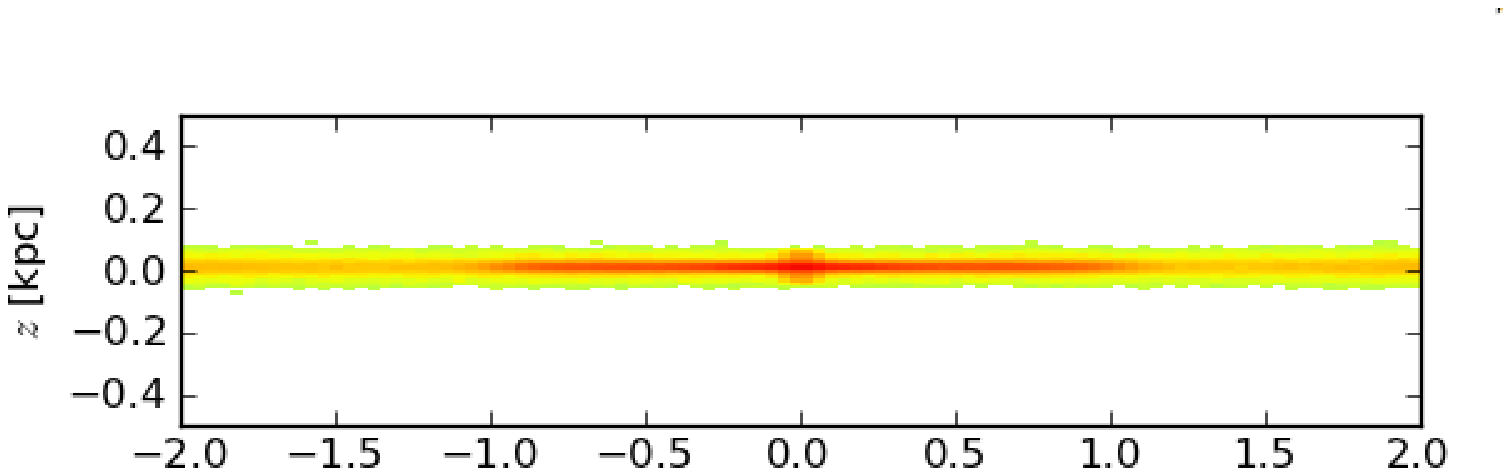}
\includegraphics[scale=0.4]{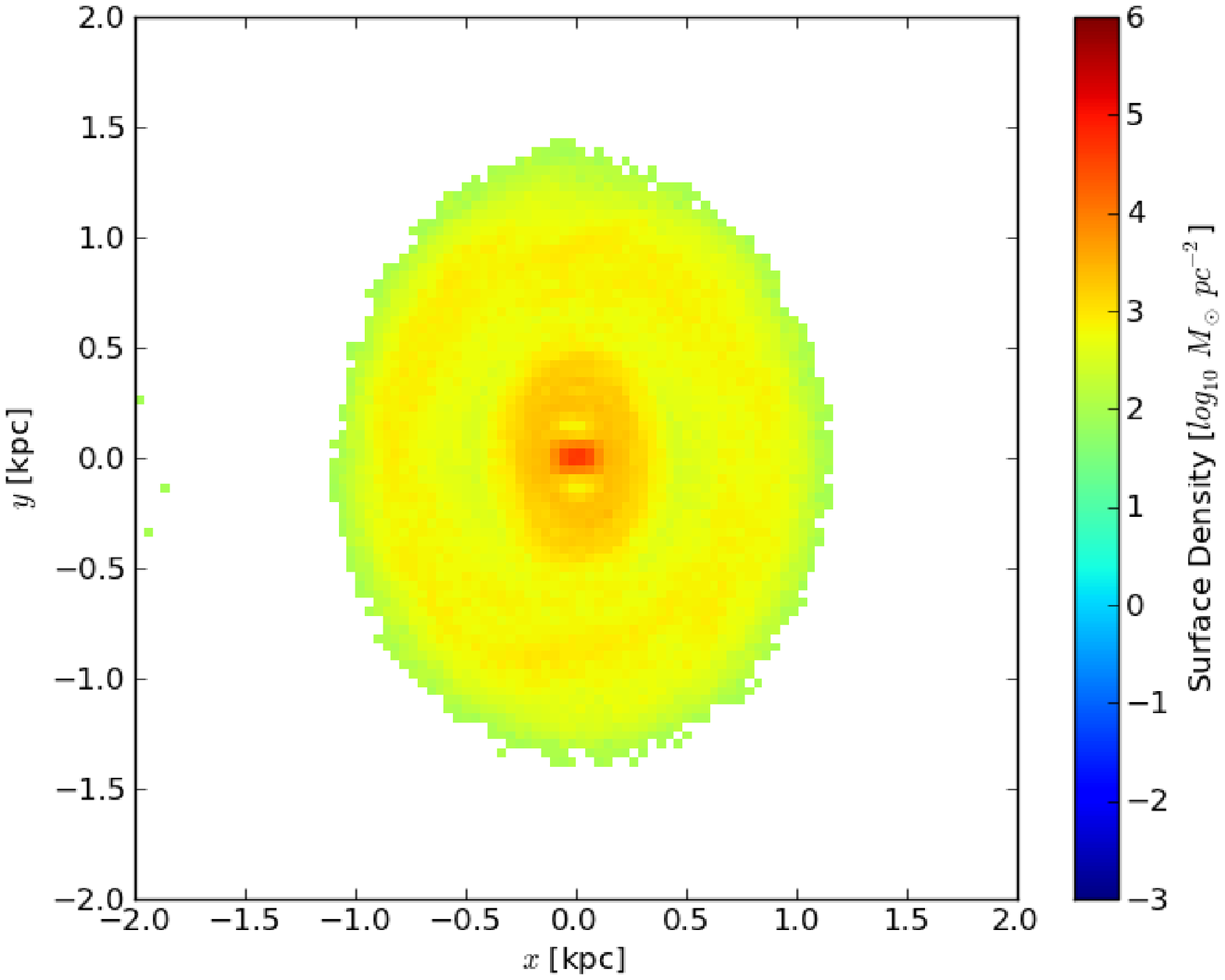}
%
%
\caption{ Face-on (bottom) and edge-on (top) stellar surface density
  for young stars ($<$2 Gyr) after 10 Gyr in the simulation of an
  $L^*$ galaxy. The galaxy has a bar which is oriented along the
  x-axis. A thin disc of stars can clearly be seen. Figure 4 of
  \cite{Cole2014}}
\label{fig:C14}       
\end{figure}

\cite{Cole2014} presented a simulation of the formation of an $L^*$
isolated galaxy.  After the bar formed, a ND developed (see Figure
\ref{fig:C14}). They demonstrated that gas flows to the centre and
fuels star formation. The resulting ND is elongated perpendicular to
the main bar, suggesting that the stars in the ND are on x2 orbits.
The ND can clearly be seen in the kinematics and the stellar
metallicity.

Given the available data, ND formation through dissipationless
processes cannot be excluded.  \cite{Agarwal2011} proposed that NDs
form out of the debris of infalling star clusters and
\cite{Portaluri2013} showed that such a scenario is consistent with
the available kinematic and photometric data. It has also been
demonstrated that NDs can be formed from accreted dwarf satellites
settling into rotationally supported NDs \citep{Eliche-Moral2011}.

\section{The Link to Pseudobulges}

Just like NSCs and NDs, pseudobulges require the inflow of gas to
form.  Pseudobulges have proven to be very common, with
\cite{Fisher2011} estimating that they account for $\sim 80\%$ of disc
galaxies.  \cite{Kormendy2004} reviewed the formation of pseudobulges
via the funnelling of gas through non-axisymmetric structures, such as
bars.  They list eight key properties of NSCs which need to be
understood if NSCs and pseudobulges are related secular phenomena: (1)
NSCs are common (2) NSCs are rare in irregulars (3) NSCs are fairly
homogeneous in their properties (4) NSCs are at the centres of their
host galaxies (5) NSCs host young stars (6) NSCs are not more common
in barred galaxies (7) In the Fundamental Plane NSCs are more similar
to globular clusters and (8) The masses of NSCs correlate with the
luminosities of their host galaxies.  They argue that points (2), (3),
(6) and (7) appear inconsistent with NSCs and pseudobulges being
related phenomena.  Possible answers to these problems could be that
the centres of irregulars are not well-defined (point 2), that
globular cluster mergers are responsible for part of the mass assembly
of NSCs (points 3 and 7) and the gas flows required for NSCs are not
as large as needed for pseudobulges (point 6).  Nonetheless, NSCs {\bf
  are} different.  \cite{Walcher2005} tested the idea that NSCs are
proto-bulges which form through in situ star formation but whose
growth has not been sufficient to form a bulge.  On the basis of 9
bulgeless galaxies, they showed that the dynamical properties of the
NSCs are very different to those bulges.

How can gas get funnelled to NSCs?  \cite{Kormendy2004} invoked bars
and ovals to explain the formation of pseudobulges.  It has long been
recognised that, while bars can drive gas inwards, this gas stalls at
the inner Lindblad resonance.  \cite{Shlosman1989} proposed that gas
can be driven all the way to the centre of a galaxy, thereby feeding
AGN activity, by means of nested bars, where a small-scale bar resides
inside a larger bar.  Such double-barred galaxies have been observed
in about $25\%$ of early-type galaxies \citep{Erwin2002}.  In this
scenario the main bar of a galaxy would induce an inward flow creating
a nuclear gas disc which could again become unstable leading to
further gas infall.  Evidence for gas inflow that can be explained by
this scenario comes from the molecular gas in NGC~6946
\citep{Schinnerer2006, Schinnerer2007}, which appears to be streaming
along the leading edge of an inner stellar bar about 400 pc long
nested inside a large-scale (3.5 kpc) bar.

However, the fact that NSCs do not prefer barred galaxies
(\citep{Carollo2002, Boker2004}) suggests that bars are not the sole
mechanism responsible for funnelling gas to nuclei.  As an alternative,
\cite{Milosavljevic2004} proposed that the magneto-rotational
instability could transport neutral gas inside 100 pc where it could
form stars.

NDs can sometimes be directly associated with pseudobulges through the
phenomenon of $\sigma$-drops, where galaxies have a significant drop
in velocity dispersion in their centre \cite[e.g.][]{Emsellem2001}.
These can be explained by infalling gas forming a dynamically cool ND.
Star formation reduces the central velocity dispersion
\citep{Wozniak2003, Comeron2008}. Small NDs have been observed with
      {\it HST} in the centre of galaxies co-located with
      $\sigma$-drops \citep{MendezAbreu2014}.

\section{Co-Evolution of SMBHs and NSCs}
\label{subsec:nucdyn}

A small fraction of galaxies host both a NSC and a SMBH
\citep{Seth2008b}, although the actual fraction could be higher given
the difficulties in detecting both in a given galaxy.
\cite{Neumayer2012} noted that a plot of M$_{bh}$ versus M$_{NSC}$
divides into three regions, one which is NSC dominated, a transition
region and one which is SMBH dominated.  This led them to speculate
that SMBHs form inside NSCs but outgrow and destroy them when the NSC
mass is less than one percent of the SMBH mass.  Alternatively,
\cite{Nayakshin2009} proposed a competitive feedback to explain the
dichotomy between NSCs and SMBHs.  Arguing that NSC growth depends on
the dynamical time of the nuclear region, they find that there is a
transition when the velocity dispersion of the host spheroid is $\sim
150$ km s$^{-1}$. Above this the NSC cannot grow efficiently and below
this the SMBH cannot grow efficiently thus explaining why NSCs are
mainly found in low and intermediate mass galaxies.
\cite{Antonini2012}, and \cite{Antonini2013} explained this dichotomy
in the globular cluster formation scenario in the presence of a SMBH.
For a low mass SMBH, such as the Milky Way's, globular clusters manage
to reach the centre allowing the NSC to grow.  However, if M$_{\rm bh}
\sim 10^8$ M$_\odot$ then the globular clusters are disrupted before
reaching the nucleus.  

However \cite{Kormendy2013} point out that there is no segregation
into giants that only contain SMBHs or dwarfs that only contain
nuclei. Where SMBHs and NSCs co-exist the ratio of SMBH to NSC mass
can vary across a large, and {\it apparently continuous} range above
and below unity.  For example in NGC~4026
$\frac{M_{bh}}{M_{NSC}}=12.4$ \citep{Lauer2005} whereas in the Milky
Way $\frac{M_{bh}}{M_{NSC}}=0.15\pm0.075$ \citep{Launhardt2002}.

Over long timescales, \cite{Merritt2009} showed that NSCs evolve under
two competing processes, core collapse by two body interactions, and
heating from the surrounding galaxy.  Which of these two processes
wins out depends on the concentration of the NSC \citep{Quinlan1996}.
However, the presence of a SMBH inhibits core collapse implying that
in this case a NSC can expand for ever and ultimately be disrupted.

\section{Scaling Relations}
\label{sec:scarels}

The masses of SMBHs, M$_{bh}$ are well known to correlate with their
host galaxy properties including the bulge velocity dispersion,
$\sigma_e$ \citep{Ferrarese2000, Gebhardt2000, Merritt2001,
  Tremaine2002, Ferrarese2005, Gultekin2009, Graham2011,
  McConnell2011, Beifiori2012}, the bulge mass, M$_{bul}$
\citep{Magorrian1998, Marconi2003, Haring2004, Sani2011, Beifiori2012,
  Graham2012} and the bulge luminosity, L$_{bul}$ \citep{Kormendy1995,
  McLure2002, Marconi2003, Graham2007, Gultekin2009, Sani2011,
  McConnell2011, Beifiori2012, Graham2013}. (See Chapter 5 of this
volume ``Galaxy bulges and their massive black holes'' by Alister W.
Graham for an up-to-date discussion on SMBH scaling relations and the
review by \cite{Kormendy2013}.)  Similarly the luminosity and mass of
nuclear star clusters and nuclear discs have been found to correlate
with their host galaxy properties \citep{Balcells2003, Balcells2007,
  Graham2003, Ferrarese2006, Wehner2006, Graham2012} which has led to
them being lumped along with SMBHs as a generic class of objects, the
central massive objects (CMOs), with stellar CMOs being found in less
massive galaxies.

\begin{figure}[b]
\sidecaption
\includegraphics[scale=2.3]{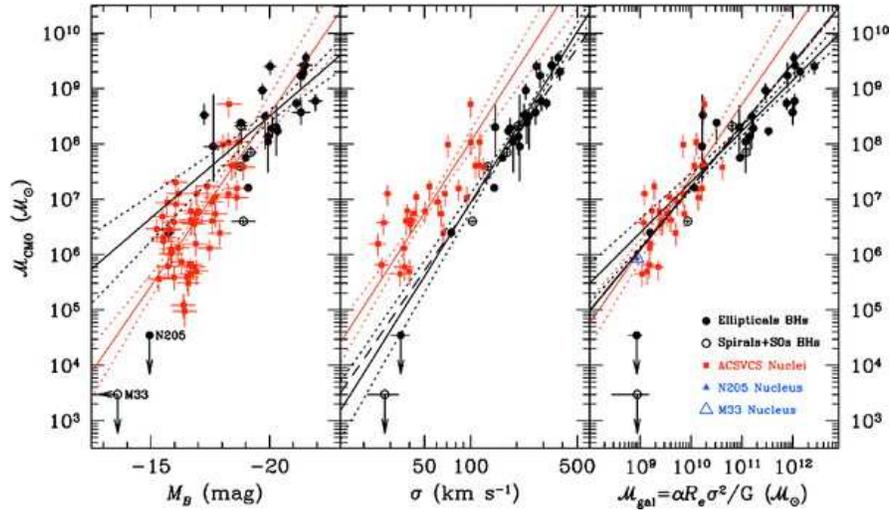}
\caption{ The mass of the CMO versus the blue band magnitude (left),
  velocity dispersion $\sigma$ (middle) and host spheroid dynamical
  mass (right) for SMBHs (filled and open circles) and nuclei (red
  squares). The solid red and black lines show the best fits to the
  nuclei and early-type SMBH samples, respectively. In the right
  panel, the dashed line is the fit obtained for the combined
  nuclei+SMBH sample. Figure 2 of \cite{Ferrarese2006}}
\label{fig:F06} 
\end{figure}

The luminosity of stellar CMOs was found to correlate with that of
their host bulge in disc galaxies by \citep{Balcells2003, Balcells2007}
and a similar correlation was found in a study of dE galaxies in the
Coma cluster by \citep{Graham2003}. \cite{Ferrarese2006} found that the
mass of stellar CMOs in early-type galaxies correlates with both the
host galaxy's luminosity and dynamical mass,
M$^{gal}_{dyn}\propto\sigma^2_eR^{gal}_e$ (Figure \ref{fig:F06}). More
importantly they found a common M$_{CMO}$-M$_{gal}$ relationship for
galaxies with either NSCs or SMBHs, with NSCs occupying fainter
galaxies with lower $\sigma_e$.  This common relation suggests that
there is a single mechanism responsible for regulating the growth of
CMOs.  They speculated that stellar nuclei form in all galaxies but in
the most massive ones they collapse to a SMBH. A similar common
relation between CMO mass and the mass of the host galaxy was also
found by \cite{Wehner2006} in dwarf elliptical galaxies.
\cite{McLaughlin2006} proposed that the mechanism responsible for this
correlation is momentum-driven feedback, from supernovae in the case
of NSCs and from AGN activity in the case of SMBHs.

\begin{figure}[b]
\sidecaption
\includegraphics[scale=5.5]{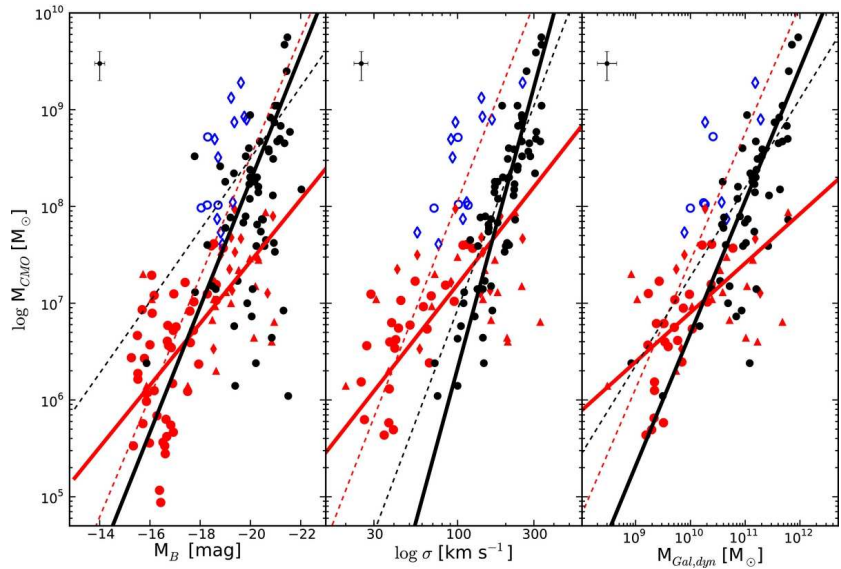}
\caption{The mass of the CMO versus the B band magnitude, velocity
  dispersion $\sigma$ and galaxy dynamical mass for SMBHs (black
  dots), NDs (open symbols) and NSCs (red dots). The solid black lines
  shows the slope of the relations for SMBHs and the solid red lines
  for NSCs. Figure 2 of \cite{Scott2013a}.  The scaling relation for
  SMBHs is updated in Figure 2 of \cite{Graham2013}, and Figure 3 of
  \cite{Scott2013b}.}
\label{fig:SG13}  
\end{figure}

However the existence of scaling relations in common between NSCs and
SMBHs has recently been questioned.  The first indication that stellar
CMOs and SMBHs do not, in fact, follow the same scaling relations came
in a study of S0-Sbc galaxies by \cite{Balcells2007}. They found that
the near infra-red luminosities of NSCs scale with host bulge
luminosities. However, in contrast to \cite{Ferrarese2006}, when they
added SMBHs they found a nonlinear dependence between $M_{\rm CMO}$
and $M_{\rm bulge}$.  An expanded dataset allowed \cite{Graham2012} to
show that NSC mass correlates with host spheroid velocity dispersion
as $log[M_{NSC}/M_\odot]=1.57\pm0.24$
log$[\sigma/70kms^{-1}]+(6.83\pm0.07)$. The slope of this relation,
$\sim2$, is much lower than that for SMBHs, $\sim 5$.
\cite{Leigh2012} found that NSC mass is directly proportional to host
spheroid mass; the virial theorem then implies an M$_{NSC}$-$\sigma_e$
relation with a slope again close to 2.  Figure \ref{fig:SG13}, taken
from \cite{Scott2013a}, shows their relations between CMO mass versus
galaxy magnitude, $\sigma$ and galaxy virial mass; these relations can
be compared directly with those of \cite{Ferrarese2006}, shown in
Figure \ref{fig:F06} find a slope of $2.1 \pm 0.3$ for the
$M_{NSC}-\sigma$ relation, much shallower than the relation found by
\cite{Ferrarese2006}. A major reason for this difference is their
inclusion of NSCs in more massive galaxies and the exclusion of NDs.
\cite{Erwin2012} and \cite{Scott2013a} both noted that the mass of
NSCs correlates better with the host's total stellar mass, whereas
that of SMBHs correlates better with the host spheroid. They conclude
that different physical processes regulate NSC and SMBH growth.

\cite{Kormendy2013} reach a more nuanced conclusion on the relation
between SMBHs and NSCs by taking galaxy type into account when
studying the ratio of CMO to bulge or galaxy mass.  NSCs in spheroidal
galaxies are relatively more massive than in late-type galaxies,
consistent with the generally held view that spheroidal galaxies are
late-type galaxies which have lost baryonic mass.  This renders these
galaxies less useful for comparing CMO scaling relations.  They find
that the ratio $(M_{bh}+M_{NSC})/M_{bulge}$ has less scatter than
either $M_{bh}/M_{bulge}$ or $M_{NSC}/M_{bulge}$, suggesting that the
evolution of NSCs and SMBHs is tightly coupled.  As a fraction of
total galaxy mass, instead, both SMBHs and NSCs have a larger relative
mass in early-type galaxies compared with late-type galaxies
(excluding the spheroidals).  \cite{Kormendy2013} conclude that this
hints at SMBHs and NSCs being related.

Although a common scaling relation between NSCs and SMBHs now seems
dead, nonetheless the existence of scaling relations between NSCs and
their host galaxies still provide important constraints on how their
growth is regulated \citep{Silk1998, King2003, Wyithe2003,
  DiMatteo2005, Murray2005, Sazonov2005, Younger2008, Booth2009,
  Johansson2009, Power2011, Wyithe2003, DiMatteo2005, Murray2005,
  Sazonov2005, Younger2008, Johansson2009}.  Theoretical models of
formation mechanisms make predictions for scaling relations between
NSCs and host galaxy properties, based on underlying physics, and
these allow us to distinguish how NSCs are formed.  The main process
proposed is feedback from the CMO, and possibly the effect this has on
its galaxy but the exact mechanisms is not yet clear \citep{Silk1998,
  Springel2005, Booth2009, Fabian1999, King2003, King2005, Murray2005,
  McLaughlin2006, King2010, Power2011, McQuillin2012}.

\section{Conclusions}
\label{sec:concs}

Observations of nuclear star clusters show that these systems can
unravel the mass assembly at the centres of galaxies. Their properties
give us clues as to how they were formed and the physical processes
that contributed to their formation. There is a close connection
between the the formation of NSCs and their host, as is demonstrated
by their scaling relations. Observations support both in situ star
formation and globular cluster merging for the formation and growth of
NSCs. How much these mechanisms contribute to the growth of NSCs
probably depends on whether they are found in early or late-type
galaxies or in high-mass or low-mass galaxies.  It also seems likely
that the morphology of the bulge, whether it is a classical bulge or a
discy bulge formed through secular processes, will affect the
transport of gas to the nuclear regions of a galaxy where it can form
stars. NSCs in early-type galaxies with little gas are unlikely to
grow due to dissipational processes. However late-type galaxies show
multiple stellar populations with stars of the order of a few hundred
Myr old implying recent star formation.

NSCs and SMBHs can co-exist, as can be seen in the Milky Way, but they
no longer seem to be two types of a single central massive object.
However studying the interrelationship between these two types of
nuclear system will contribute to the understanding of all physical
processes which are important in their formation. Further
investigation of how scaling relations are affected by the presence of
a bar, the morphology of the bulge and whether the host galaxy is
early- or late-type may allow us to refine our ideas of how they form.
There is also a need for higher resolution simulations of the effects
of in situ star formation on the kinematics, chemistry and morphology
of NSCs. From the information we have gathered so far it is clear that
NSC formation is very complicated and their continued study will bring
further insights.
\\
\\
\textbf{Acknowledgements}\\
DRC and VPD are supported by STFC Consolidated grant ST/J001341/1.

\bibliographystyle{mn2e1}
\bibliography{NSC}

\end{document}